\documentclass[sigconf, nonacm]{acmart}
\AtBeginDocument{%
  }
\usepackage{balance} 
\usepackage{algorithm}
\usepackage{algorithmic}

\begin{document}

\title{ZV-Sim: Probabilistic Simulation Framework for Pre-emergent Novel Zoonose Tracking}

\author{Joseph Maffetone}
\affiliation{%
  \institution{University of Michigan}
  \city{Ann Arbor}
  \country{Michigan}}
\email{jmaff@umich.edu}

\author{Julia Gersey}
\affiliation{%
  \institution{University of Michigan}
  \city{Ann Arbor}
  \country{Michigan}}
\email{gersey@umich.edu}

\author{Pei Zhang}
\affiliation{%
  \institution{University of Michigan}
  \city{Ann Arbor}
  \country{Michigan}}
\email{peizhang@umich.edu}

\renewcommand{\shortauthors}{Maffeton et. al}

\begin{abstract}
ZV-Sim is an open-source, modular Python framework for probabilistic simulation and analysis of pre-emergent novel zoonotic diseases using pervasive sensing data. It incorporates customizable Human and Animal Presence agents that leverage known and simulated location data, contact networks, and illness reports to assess and predict disease origins and spread. The framework supports Monte Carlo experiments to analyze outcomes with various user-defined movement and probability models. Although initial models are basic and illustrative, ZV-Sim's extensible design facilitates the integration of more sophisticated models as richer data become available, enhancing future capabilities in zoonotic disease tracking. The source code is publicly available \href{https://github.com/jmaff/zv-sim}{\underline{\textit{here}}}. 

\end{abstract}
\maketitle

\section{Introduction}

Emergent zoonotic diseases, such as SARS, MERS, and COVID-19, have repeatedly demonstrated their potential for devastating impact once sustained human-to-human transmission is achieved \cite{clemente-suarez_impact_2021, rahman_zoonotic_2020}. Early identification of zoonotic spillover events is thus critical to enable early intervention, but remains an open problem due to the complex interaction surfaces between animals and humans, as well as the lack of cohesive methods to integrate data relevant to tracking such interactions \cite{yin_survey_2025}.

Traditional epidemic modeling frameworks focus primarily on human mobility and human-human transmission dynamics, offering limited ability to model the initial stages of zoonotic spillover where human-animal interaction plays a critical role \cite{yin_survey_2025, barrett_episimdemics_2008, plowright_pathways_2017, ellwanger_zoonotic_2021}. While some recent efforts have incorporated animal movement data to estimate risk and understand spread dynamics, these works are typically rigidly focused on a certain disease and animal population, and do not provide freely available frameworks that could be adapted to study emerging disease \cite{gilbert_predicting_2014}. Moreover, many existing simulations lack the ability to integrate both known and simulated sensing data, which could improve early inference about the source of infections \cite{de_angelis_four_2015, witt_simulation-based_2020}. In the context of growing availability of pervasive sensing technologies, such as GPS-based location tracking and animal migration monitoring, there is an opportunity to create a simulation framework that can integrate diverse, noisy, and partially observed data sources to monitor for novel zoonoses before widespread transmission occurs. 

As such, we introduce ZV-Sim, an open-source, modular simulation framework for modeling zoonotic disease emergence. In this paper, each component of the framework is detailed, including data sources and animal/human agents, core simulation logic, user-defined models and extensions, and outputs. Last, the powerful analysis enabled by ZV-Sim is demonstrated via experiments using simple yet illustrative models and datasets.

\begin{figure}[ht]
    \centering
    \includegraphics[width=0.43\textwidth]{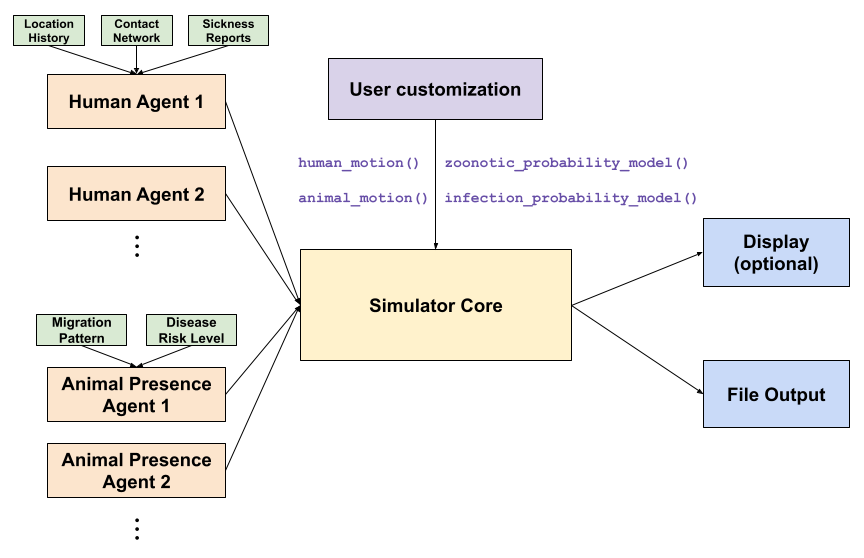}
    \caption{ZV-Sim Architecture}
    \label{fig:blockdiagram}
\end{figure}

\section{High-Level Architecture}
ZV-Sim has a modular organization, allowing for minimal coupling between input, simulation logic, and output. This in turn allows modifications to individual modules without the need to extensively modify the entire simulator codebase, improving accessibility for future work to extend functionality. A block diagram of these modules is shown in Figure~\ref{fig:blockdiagram}, and each component is covered in detail in Sections 3-5. 

\section{Agents}
Agentic simulation has shown promise in high accuracy for both animal and human behaviors and motion \cite{yin_survey_2025,tang_agent-based_2010, scharf_animal_2020, milner_modelling_2021}. For this reason, ZV-Sim uses agents to model and simulate both animals and humans in the context of disease tracking. 

\subsection{Human Agents}

\subsubsection{Inputs}
Human agents are initialized with the following input data:

\begin{description}
  \item[\textbf{ID:}] A unique numerical identifier for the human agent
    
  \item[\textbf{Location History:}] Known, timestamped records, mapped to the simulation grid

  \item[\textbf{Sickness Reports:}] Known, timestamped reports (sick or healthy), such as self-reports or medical records
\end{description}

Location history is the primary input for human agents, and was selected due to being a strong datapoint for monitoring disease spread \cite{ghayvat_recognizing_2021, zhang_simulating_2022, grubaugh_tracking_2019}. As discussed in Section~\ref{sec:core}, human location history is used to build contact networks, calculate animal and human disease exposure, and other factors relevant to tracking disease spread and origin.  

\subsubsection{Internal State \& Outputs}
Throughout a given simulation run, human agents update their internal state per simulation timestep (``tick"), and ultimately store useful data for the user once the run has completed. The currently available fields are:

\begin{description}
  \item[\textbf{Status:}] Whether the human is currently healthy or sick

  \item[\textbf{Location:}] The current location of the agent (simulation space)

  \item[\textbf{Contact Network:}] List of contact records (other human's ID, time range of contact, average proximity)

  \item[\textbf{Infection Model:}] User-defined container with values relevant for disease spread modeling 

  \item[\textbf{Sickness Records:}] List of sicknesses experienced by the human agent during the simulation. This includes the number of estimated secondary cases and the ultimate probability that this illness was of zoonotic origin. 
  
\end{description}

Sickness records are the primary useful output for a human agent, as they directly include all information related to the spread and origin of disease, and as such are the result of focus in this paper. Other fields remain stored for each agent at the end of simulation for further analysis if desired. 

\subsection{Animal Presence Agents}

Due to the much higher difficulty in collecting sufficient data at an individual level for wild animals \cite{zhang_hardware_2004, hughey_challenges_2018}, animals are grouped into populations. These ``Animal Presence" agents represent a geographical area where a certain animal population is likely located. Zoonotic spillover typically requires contact with biological material from vector animals \cite{yin_survey_2025, plowright_pathways_2017}, and so Animal Presence agents include both the region where humans may contact animals directly, as well as where humans may contact their biological matter (feces, meat, blood, etc.). 

\subsubsection{Inputs}
Animal Presence agents are initialized with the following input data:

\begin{description}
  \item[\textbf{ID:}] A unique numerical identifier for the animal presence agent

  \item[\textbf{Migration Pattern:}] Timestamped records of the estimated central presence of the animal population

  \item[\textbf{Radius:}] Distance from each location record that will be considered part of the animal presence

  \item[\textbf{Hazard Rate:}] Numeric value representing relative risk of transmissible disease from the animal population (see Section~\ref{sec:spread}). 
\end{description}

Currently, animal presence agents assume a constant disease risk throughout the simulation, and only allow for circular presence regions. These limitations were primarily selected for the sake of simplicity in the initial implementation. Extensions may be made to allow these agents to mutate hazard rate according to simulated events, have more advanced geometry, or even simulate individual animals instead of populations. 

\subsubsection{Internal State \& Outputs}
 Similar to human agents, Animal Presence agents update their internal state at each simulation timestep. These fields include:

\begin{description}
  \item[\textbf{Location:}] Current central location of the animal population (simulation space)

  \item[\textbf{Infection Model:}] User-defined container also used in human agents
\end{description}

Animal presence agents notably track less internal state and have less useful output than human agents. This is because most of the animal contact information that is relevant to disease tracking is already captured by human agents. If a study wishes to focus on disease spread from the perspective of animal populations, modifications can be made to track additional records similar to human agents. 

\section{Simulator Core \& User Customization}
\label{sec:core}

The Simulator Core module is intended to require the least modification for users of ZV-Sim, relying on abstraction to easily integrate customizations to agents and models. A simulation occurs in ``simulation space", a continuous 2D grid of size \texttt{GRID\_WIDTH} $\times$ \texttt{GRID\_HEIGHT}. Time is discretized into timesteps (``ticks"), which each equate to \texttt{SIM\_TICK\_TIME\_SECONDS} real-world seconds. All of these parameters, located in \texttt{simulator.py}, can and should be customized by the user when loading data into the simulator.

The simulator core then runs \texttt{NUM\_TRIALS} trials, calculating and recording output values for each trial to file for later analysis. Several components in the trial algorithm pull directly from user-defined code, present in \texttt{user.py}, allowing for easy customization of models used in simulation. These customizable models are detailed in the following subsections, along with the included implementations primarily intended for demonstration.

\subsection{Motion Models}
\label{sec:motion}

Location data is necessarily collected at discrete timestamps. In many common sensing scenarios, data sampling rates may be further reduced to conserve power, storage, bandwidth, etc \cite{clark_advanced_2006, paek_energy-efficient_2010}. As such, in a given simulation run there will likely be many timesteps without known location data for a given agent. Since most disease-relevant metrics available from the core simulation module rely on location data (contact networks, hazard levels, etc.), assuming a static location at these ticks may hinder simulation accuracy. Motion models are invoked to predict a new location for an agent when known location data isn't available, and receive the current and next known locations as inputs. Separate models can be defined for animal and human motion, and are fully modifiable in the \texttt{user.py} file. 

Besides choosing to only use known location data (``None"), other simple motion models have been included in ZV-Sim, and are outlined below.

\subsubsection{Random Walk}
In the random walk motion model, future known locations are ignored entirely. Instead, it is assumed that agents will remain relatively close to their current locations, with some additive noise. This noise is represented by two uniformly distributed, bounded random variables that are added to the agent's previous X and Y locations, respectively. The model is formalized below, with an example range of $\{-5, \ldots, 5\}$ for all noise. 
\begin{align*}
x_t &= x_{t-1} + \epsilon_x, \quad \epsilon_x \sim \mathcal{U}(\{-5, \ldots, 5\}) \\
y_t &= y_{t-1} + \epsilon_y, \quad \epsilon_y \sim \mathcal{U}(\{-5, \ldots, 5\})
\end{align*}

\subsubsection{Noisy Linear Interpolation}
This model assumes a linear path from the last known location of an agent to its next known location, with some additive noise along the path. While still quite simple, this approach allows the simulator to effectively ``fill in" missing location data, which may capture possible interactions that otherwise would be missed. The model is described below, using uniform distributions for noise similar to the Random Walk model.

\begin{gather*}
\text{Let $d$ be the next timestep for which the agent has a known location.} \\
\begin{aligned}
\Delta x = x_d - x_{t-1} &&
\Delta y = y_d - y_{t-1} &&
\Delta t = d - (t-1) 
\end{aligned}
\\
\begin{aligned}
x_t &= x_{t-1} + \frac{\Delta x}{\Delta t} +\epsilon_x, \quad \epsilon_x \sim \mathcal{U}(\{-5, \ldots, 5\}) \\
y_t &= y_{t-1} + \frac{\Delta y}{\Delta t} +\epsilon_y, \quad \epsilon_y \sim \mathcal{U}(\{-5, \ldots, 5\})
\end{aligned}
\end{gather*}

\subsection{Disease Spread Model}
\label{sec:spread}
Reported illness is not a completely reliable metric for tracking (novel) disease, as many sick individuals likely will not seek medical attention, and relevant records are difficult to obtain and organize even when available \cite{hogan_accuracy_1997, gibbons_measuring_2014}. As such, ZV-Sim has the capability to simulate sickness of human agents outside of reported data, according to the user-defined Disease Spread Model. 

The Disease Spread Model defines both a function to calculate the probability that a given human agent has become ill at each timestep (\texttt{infection\_probability\_model}), as well as the \texttt{InfectionModel} dataclass that is persisted as state for each agent, containing fields needed by the model. 

\subsubsection{Hazard Model}
ZV-Sim relies heavily on location data and simulation, and thus primarily models agent \textit{exposure} to possible disease vectors. To translate this exposure into probability of disease transmission, the well-established cumulative hazard model (survival analysis) \cite{ohagan_estimating_2014, selvin_survival_2008} was implemented. We model disease vectors (sick humans, at-risk animal presences, etc) with some numerical hazard that is accumulated by human agents per timestep they are exposed to the vector. Based on this accumulated hazard, $\lambda$, the probability that a given agent will become ill is equal to the CDF of an exponential random variable:
\[
P(\text{infection}) = 1 - e^{-\lambda}
\]
Separate fields for accumulated hazard from animal and human vectors are recorded for use in models concerning disease origin, but are summed when considering disease spread here. Hazard values for various vectors are estimated and rudimentary, but can and should be replaced by data-informed values. 

\subsection{Zoonotic Probability Model}

ZV-Sim's most useful output as related to novel zoonotic disease tracking is the ability to predict whether a given case of human illness is of novel zoonotic origin. Such an ouput is represented by a probability, as calculated by the user-defined Zoonotic Probability Model. This model receives a human agent's sickness record as input, which includes metrics from simulation such as number of secondary cases, hazard levels experienced at the time of infection, etc.

An illustrative Bayesian-based model is likewise included with ZV-Sim, although incorporation of a more sophisticated approach is strongly suggested.

\subsubsection{Bayesian Zoonotic Probability Model}

Bayesian probability provides a framework for updating beliefs in the presence of new evidence, making it well-suited for modeling an uncertain process like disease origin estimation \cite{andradottir_applying_2000, mckinley_approximate_2018, mckinley_simulation-based_2014}. ZV-Sim includes a Bayesian model that combines prior knowledge about zoonotic disease risk with simulator data for a given sickness record, namely animal exposure at the time of infection and the number of likely secondary cases. Naturally, higher hazardous animal exposure at time of illness indicates a higher probability of zoonotic origin. Additionally, a \textit{low} number of secondary cases also indicates higher probability of a novel zoonose, as effective human-human transmission is typically not present at initial spillover \cite{yin_survey_2025, saldana_modeling_2024, plowright_pathways_2017, ellwanger_zoonotic_2021}.

We begin by defining the binary random variable $O$ to denote the origin of an observed illness, with possible values Z and H to denote ZOONOTIC and HUMAN origin, respectively.
% \begin{gather*}
% O \in \{\text{Z},\,\text{H}\} \\
% P\bigl(O = \text{Z}\bigr) = p,
% \qquad
% P\bigl(O = \text{H}\bigr) = 1 - p.
% \end{gather*}

Additionally, we define as inputs to our model the total experienced animal hazard ($\lambda_A$), total number of likely secondary cases ($k$), and the prior probability of zoonotic infection ($Z_p$).

This allows us to then define: 
\[
\begin{aligned}
  &P(\lambda_A \mid O=\mathrm{Z}) = 1 - e^{-\lambda_A}
  &&\text{(cumulative hazard model)} \\[1ex]
  &P(\lambda_A \mid O=\mathrm{H}) = 1 - P(\lambda_A \mid O=\mathrm{Z})
  &&\text{(assumption)} \\[1ex]
  &P(k\mid O=\mathrm{Z}) = \frac{0.1^k e^{-0.1}}{k!} 
  &&\text{(Poisson distribution)} \\[1ex]
  &P(k\mid O=\mathrm{H}) = \frac{2^k e^{-2}}{k!} 
  &&\text{(Poisson distribution)} 
\end{aligned}
\]

The PMF for secondary case likelihood is centered around a higher mean when human origin is given than zoonotic origin (2 expected cases versus 0.1). Of note is that the assumption made to calculate $P(\lambda_A \mid O=\mathrm{H})$ may be ill-formed, as $\lambda_A$ is continuous, and is a limitation of this included model that should be addressed in future work. 

Last, we use Bayes' theorem to derive the probability that the origin of the illness is zoonotic ($O = \mathrm{ZOONOTIC}$) \textit{given} the data available from simulation ($\lambda_A$ and $k$). 

\begin{align*}
&P(O=\mathrm{Z} \mid \lambda_A, k) = \\
&\frac{P(\lambda_A \mid O=\mathrm{Z}) \, P(k\mid O=\mathrm{Z}) \, Z_p}{P(\lambda_A \mid O=\mathrm{Z}) \, P(k\mid O=\mathrm{Z}) \, Z_p + P(\lambda_A \mid O=\mathrm{H}) \, P(k\mid O=\mathrm{H}) \, (1 - Z_p)}
\end{align*}

This is our desired probability: the probability that a case of disease was of novel zoonotic origin given the metrics calculated by the simulator. 

\setcounter{figure}{2}
\begin{figure*}[t!]
    \centering
    \includegraphics[width=0.91\textwidth]{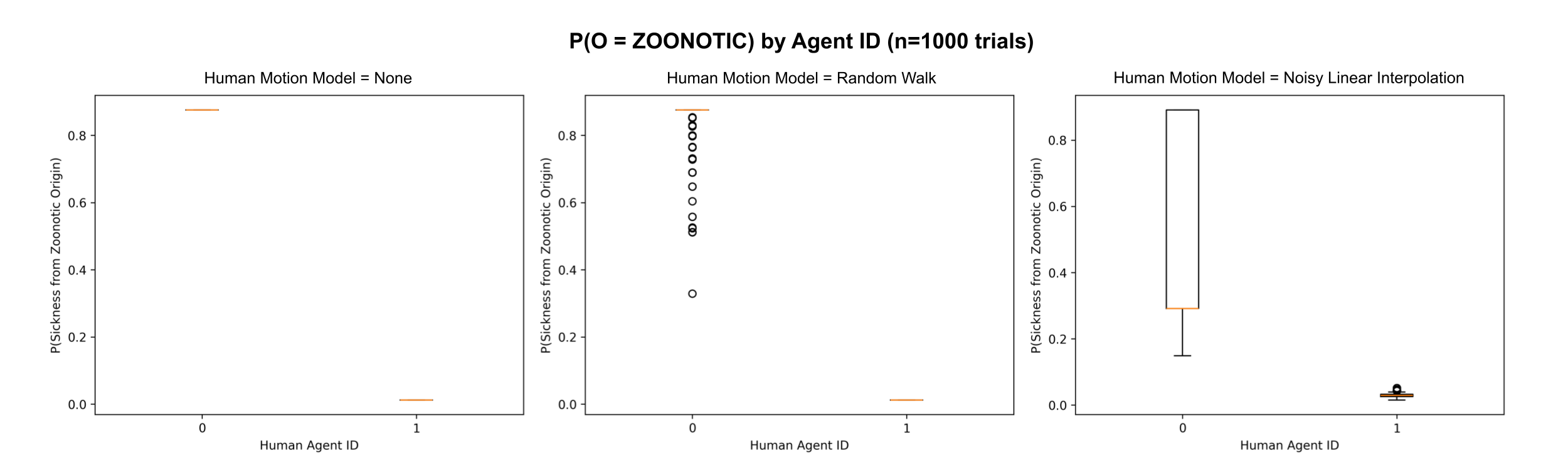}
    \caption{Probability of novel zoonotic origin for the illnesses experienced by each human agent, compared across motion models None (left), Random Walk (center), and Noisy Linear Interpolation (right)}
    \label{fig:results}
\end{figure*}

\setcounter{figure}{1}
\begin{figure}[ht]
    \centering
    \includegraphics[width=0.3\textwidth]{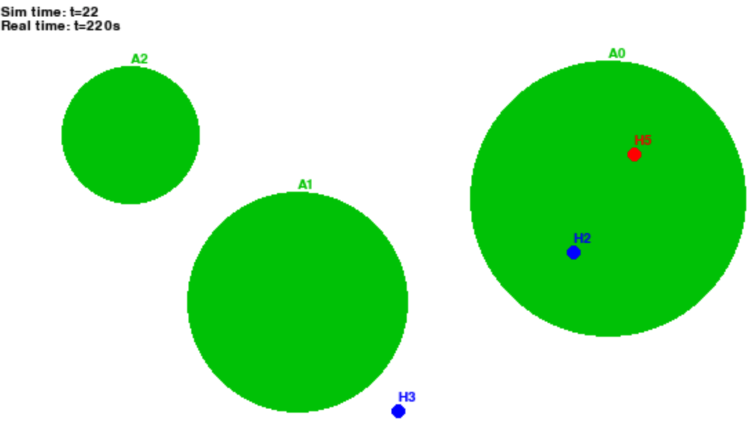}
    \caption{Included PyGame display module implementation. Human agents are colored red if sick.}
    \label{fig:pygame}
\end{figure}

\section{Outputs}
ZV-Sim includes multiple easily customizable output methods for simulation data, for both ``real-time" visualization and off-line analysis. 

\subsection{Display Module}

The display module is responsible for rendering simulation state to some user-friendly format at each simulation timestep. The included implementation uses the PyGame \cite{noauthor_pygamepygame_nodate} framework to display one frame per timestep, representing human agents as colored dots, and animal presence agents as green circular regions, shown in Figure~\ref{fig:pygame}. Using the display module is completely optional, and not recommended when running multi-trial simulations.

% todo: include image of pygame windwo

\subsection{Monte Carlo Experiments}

Since ZV-Sim has many stochastic components, the results of a single trial may not hold much value in a real-world case study. Rather, an analysis of distributions that emerge across a large number of trials for a certain dataset would provide much greater value. ZV-Sim natively supports such Monte-Carlo experiments \cite{harrison_introduction_2010, maltezos_novel_2021}, exposing \texttt{NUM\_TRIALS} as a parameter to the user. Final sickness record values for each human agent are then stacked into a NumPy vector by trial number, which is directly saved to file for later analysis. 

\section{Results}

Since the focus of this paper is purely on illustrating the capabilities of ZV-Sim and future extensions, we create a simple small-scale dataset intended to demonstrate the analysis capabilities of the tool. This dataset consists of two human agents, $H_0$ and $H_1$, as well as two animal presence agents, $A_0$ and $A_1$. The simulation grid is $600\times 600$. Location data was written so that $H_0$ would spend much more time in contact with high-hazard $A_0$, while $H_1$ only has brief contact with low-hazard $A_1$. $H_0$ reports illness shortly after the animal contact. The human agents then move past each other, shortly after which $H_1$ becomes ill. However, $H_1$ has unreliable location data, and does not directly include the datapoints where $H_0$ is close enough for contact. 

Monte-Carlo experiments ($n=1000$ trials) were run with this dataset for three different human motion models: None, Random Walk, and Noisy Linear Interpolation (see Section~\ref{sec:motion}). Box plots for the sickness records of $H_0$ and $H_1$ across all trials by motion model are provided in Figure~\ref{fig:results}.  

As expected, $H_0$'s sickness record shows significantly higher probability of novel zoonotic origin than $H_1$ across all experiments. However, notable differences emerge depending on the human motion model. Only relying on known data (None) misses the fact that $H_1$'s illness likely was a secondary case from $H_0$, causing much higher zoonotic probability. Random Walk struggles less to represent this interaction, but still has a high median. Linear interpolation model, lastly, is  able to reliably capture the contact of the agents, as shown by its much lower median probability. 

These results, while being purely illustrative in nature, showcase the ability of ZV-Sim to supplement missing location data with user-defined models, integrate diverse data sources into an estimate of disease origin, and facilitate analysis of the impact of model differences on simulation outcome.

\section{Future Work}
The primary intent of this work is to provide an open-source, easily extensible framework for early tracking of novel zoonotic disease. As noted, the models, data, and parameters used to demonstrate the framework's capabilities are not suited for real-world studies. Future work should be done to parse real-world data sources, especially GPS data from humans and animals, into the simulator. Most critically, research should be done to develop and integrate more robust models for agent motion, disease origin probability, and simulated disease spread. The built-in experiment utilities help facilitate such studies, as the effects of different models on simulation outcomes can be easily examined. Upon the integration of more robust models, case studies can and should be performed that assess the accuracy of ZV-Sim against real-world data. 

\balance
\bibliographystyle{ACM-Reference-Format}
\bibliography{main}

\appendix

\end{document}